\documentclass[preprint,double-spaced,aps,apssymb,showpacs]{revtex4}
\usepackage{amsmath,amssymb}
\usepackage{graphicx}
\usepackage{dcolumn}
\usepackage{bm}
\usepackage{psfrag}
\usepackage{subfig}
\newcommand{\ber}{\begin{eqnarray}}
\newcommand{\eer}{\end{eqnarray}}
\newcommand{\bea}{\begin{equation}}
\newcommand{\eea}{\end{equation}}

\begin{document}

\title{Directional transport induced by elasticity and volume exclusion}

\author{A. Bhattacharyay}
\email{a.bhattacharyay@iiserpune.ac.in}
 \affiliation{Indian Institute of Science Education and Research, Pune, India}

\date{\today}

\begin{abstract}
We investigate an exactly solvable model for directional transport in 1D. The structured system, which has strong elastic interactions in its parts, explicitly demonstrates the role of volume exclusion in producing directional transport. We capture the complementary role of the elasticity and volume exclusion as the basic ingredient for showing up of broken microscopic symmetry at the scale of macroscopic motions. We compare the analytic results with the numerical simulation.      
\end{abstract}
\maketitle

Brownian Ratchet (BR) paradigm finds wide applications when it comes to understanding the driven directional transport of microscopic objects \cite{ast, rei, han, han1}. There are a host of microscopic biological systems like kinesin and myosin motor proteins etc which move on polar tracks being driven by intermittent energy supply through ATP hydrolysis \cite{vale}. Attempts have been made to understand such motions in terms of BR paradigm invoking numerous variants of the BR model \cite{ast1,oka,gei}. Not only microscopic biological transport, there are many other areas like flow through micro channels, nanopores and even in cold atoms and optical lattices where the BR concept is applied to understand the transport phenomenon \cite{han1,jone}. Brownian Motors - the term coined by H\"anggi \cite{han1} for a class of such thermally assisted moving systems - require the system be driven out of equilibrium. This actually is a prerequisite of any directed transport. It also requires the symmetry be broken in some form either internally or externally. In many of these BR models, an internal degree of freedom of the microscopic object is coupled to an external field which is responsible for global symmetry breaking \cite{cil,von,mogi,fog,nor,kum} to result in macroscopic transport. In most of these models, the transported object is treated as a particle considering its internal structure not of importance.
\par
Recently, there has been an interesting report of a variant of BR model, so-called intrinsic BR model, where the internal degree of freedom is not coupled to an external field for the sake of global symmetry breaking \cite{bro}. Rather, the internally broken symmetry at the microscopic scales shows up macroscopically in the transport of the object. The present work proposes an exactly solvable model that explicitly manifests reflection of microscopically broken symmetry in the macroscopic transport of the system. We will be considering in the following a structured object as the motor which has symmetry broken inside and would not take into account any repeating ratchet potential (external field) that helps generate global asymmetry. We will identify two important physical entities namely the elastic interaction between the parts of structured motor and volume exclusion as the essential and complementary requirements for the directed transport. While the elasticity allows the cross talk between the parts of the structured object, the volume exclusion is maintaining the locally broken symmetry intact and as a result of that the system shows macroscopic directional motion. We will also, substantiate our exact analytic results by comparing those with simulations.
\par
The present model takes into account, two particles connected by a spring and are immersed in a thermal bath characterized by a fixed temperature T. In such a situation the model looks like 
\ber\nonumber 
\dot{x_1} &=& -\alpha(x_1-x_2) +\eta_1\\
\dot{x_2} &=& -\alpha(x_2-x_1)+\eta_2.
\eer
where, $\alpha$ is the force constant of the spring connecting the particles sitting at $x_1$ and $x_2$ ($x_1>x_2$ say). The $\eta$'s are the thermal noises experienced by the system where $<\eta_i>=0$ and $<\eta_i(t_1)\eta_j(t_2)>=2T\delta_{ij}\delta(t_1-t_2)$ which is the fluctuation-dissipation relation. Note that, the $\alpha$ and the Boltzmann constant $\kappa_B$ have been absorbed by adjusting the scale of the temperature in the fluctuation-dissipation relation which is a very standard practice. 
\par
Such a system would soon attain equilibrium and one naturally cannot expect having directional transport. So, we put a drive to one of the particles which will be driving the system out of equilibrium by breaking the fluctuation-dissipation relation. Throughout this paper, we will consider an over damped dynamics of the system since this over damped regime is physically relevant in most of the biological and other cases. Now, in the overdamped regime, where there is no inertial term present, the velocity of a particle is proportional to the force acting on it. Taking this into consideration we will use a drive which is equivalent to feeding back the instantaneous velocity to one of the particles and that would invalidate the otherwise existing fluctuation dissipation relation. This not only is a very general way of driving the system out of equilibrium, but also it would enable us to solve the system exactly and calculate the average velocity and efficiency of the system in exact analytic forms. So, with this driving, our model looks like
\ber\nonumber 
\dot{x_1} &=& -\alpha(x_1-x_2) +\beta\dot{x_1}+\eta_1\\
\dot{x_2} &=& -\alpha(x_2-x_1)+\eta_2.
\eer  
where $\beta$ is the velocity feedback constant or the strength of the external drive. In what follows we will keep $-1\leq \beta \leq  1$ in order to remain safely within the over damped regime. A large forcing would eventually require the inertial term be taken into consideration which we would not like to move into at the moment. 
\par
In order to understand where this modified situation with the velocity feedback actually differs from the model in Eq.1 let us write it down in an analogous from as
\ber\nonumber 
\dot{x_1} &=& -\frac{\alpha}{1-\beta}(x_1-x_2) +\frac{\eta_1}{1+\beta}\\
\dot{x_2} &=& -\alpha(x_2-x_1)+\eta_2.
\eer
the velocity feedback term which we are considering here as a drive (which one might as well call the extra damping on the particle at $x_1$ than that in the position $x_2$) is in effect changing the local temperature of the particle at $x_1$ than that at $x_2$ resulting in a nonequilibrium situation. This nonequilibrium situation would sustain all over the time because within the scope of the model the Gaussian noise strength and the so called damping on the particles are held constant. Thus, effectively in this model the local environment of the two interacting particles are kept fixed at two different temperatures and that keeps the nonequilibrium situation prevailing for all the time. The above reasoning makes it clear why, in the present context, we prefer calling the feedback term as a drive than considering it merely a part of the damping on the particle at $x_1$.
\par
To analyze this model further, let us move to the center of mass (CM) $X=(x_1+x_2)/2$ and the internal gap $Z=(x_1-x_2)$ coordinates. The model now reduces to the form
\ber\nonumber
\dot{Z}=-\frac{\alpha(2-\beta)}{1-\beta} Z +\xi_Z \\
\dot{X}= -\frac{\alpha\beta}{2(1-\beta)} Z + \xi_X
\eer 
where, $<\xi_i>=0$, $<\xi_Z(t)\xi_Z(t^\prime)>=2TD_1\delta(t-t^\prime)$, $<\xi_X(t)\xi_X(t^\prime)>=\frac{T D_1}{2}\delta(t-t^\prime)$ and $D_1=(1+\frac{1}{(1-\beta)^2})$.

Eq.2 clearly shows that the internal coordinate $Z$ will have an equilibrium distribution 
\bea
P_{eq}(Z)=exp(-\frac{\alpha(2-\beta)}{2TD_1(1-\beta)}Z^2).
\eea
Since, the average velocity of the CM $<\dot{X}>$ is proportional to the $<Z>$, it can be easily calculated. Note that, although $Z$ has a probability distribution with zero mean the volume exclusion will prevent $Z$ being less than zero and thus we will actually have a nonzero $<Z>$ by simply integrating one one half of the distribution it has. This is exactly where we explicitly see the role of volume exclusion in producing directional transport. Thus, we arrive at
\ber\nonumber
<Z>&=& \sqrt{\frac{2TD_1(1-\beta)}{\pi\alpha (2-\beta)}}\\
<\dot X> &=& -\frac{\alpha \beta}{2(1-\beta)}<Z>=-\frac{\alpha \beta}{2(1-\beta)}\times\sqrt{\frac{2TD_1(1-\beta)}{\pi\alpha (2-\beta)}}.
\eer
Note that, at $\beta=0$ the average velocity of the system vanishes identically. This is in keeping with the fact that one cannot expect directed transport in equilibrium. Interestingly, in the above expression, we see that with a positive $\beta$ the system will move towards left i.e. in the direction from $x_1$ to $x_2$ and it changes sign with the change in sign of $\beta$. This selection of the direction of motion can easily be understood. The effective spring constant for the particle at $x_1$ being $\alpha/(1-\beta)$ is greater than $\alpha$ for a positive $\beta$. As a result the particle at $x_1$ will impinge on the particle at $x_2$ with more momentum and that will make the system move towards left. The situation will change with the change in the sign of $\beta$.
\par
A calculation of the efficiency of the system is quite straight forward. Considering the average input energy to the system is proportional to $<\dot{X}^2>$ (i.e. average instantaneous kinetic energy of the CM), we can take the effective energy output as $<\dot{X}>^2$. Now, the efficiency can be given as $\eta_{eff}=<\dot{X}>^2/<\dot{X}^2>$. The explicit calculation produces 
\bea
\eta_{eff}=\frac{1}{\frac{\pi}{2}[1+\frac{2(1-\beta)(2-\beta)}{\alpha\beta^2}]}.
\eea
Some interesting features can immediately be found out from the above relation for the efficiency. The efficiency being not dependent on the temperature it explicitly shows that the thermal energy is not being drained out. But, it is helping the system achieve directional transport because, the average velocity of the CM is a function of temperature. This explicit non-dependence of the efficiency on temperature is an important feature. The maximum efficiency (for $\beta=1$) is $100\times 2/\pi$ which is roughly $64\%$. This is a big percentage efficiency indeed. The efficiency goes linearly to zero with $\alpha$ and with $\beta$. 
\par
To compare the analytic results we simulate the system. While simulating the system, to apply the volume exclusion i.e. correcting for particle positions when they cross each other during sequential updating we work at two extreme limits namely particles suffering totally elastic and inelastic collisions. In the totally elastic collision scheme, we swap the particle positions following any violation of the volume exclusion and thus the CM remains unaltered preserving the momentum. In the totally inelastic scheme, we place the particles at the CM following any violation of the volume exclusion constraint. In Fig.1a and b we have plotted the average velocity of the system against $\alpha$ and $\beta$ respectively at some constant value of the other parameters. During these simulations we have kept the strength of the thermal noise $<\eta^2>=2 T= 0.0001$. In all the simulations we have taken the maximum positive value for $\beta$ to be 0.99 because the average velocity of the system diverges at $\beta = 1$. In Fig.2a and b we have plotted the efficiency of the system against the parameters $\alpha$ and $\beta $ as well under similar situations. In these figures a comparison of simulation results with theoretically calculated numbers is excellent and directly justifies our exact calculations.
\par
In most of the BR models actual structural realities like volume exclusion etc are masked due to considering the system as a structureless particle. Moreover, considering a repeated setup of asymmetric microscopic potentials in 1D is also in effect a macroscopic symmetry breaking that results in macroscopic directional transport. With the help of the present model, we explicitly show that in 1D there is no need for such macroscopically broken symmetry present for the transport to happen. Considering the internal structure of the microscopic moving system, we show that the volume exclusion would actually make the microscopic broken symmetry remain intact over all the time in 1D where there is no rotational degrees of freedom. Thus, the direction of motion picked up by the system will be there for ever if the nature of the instantaneous velocity feedback remains the same. In this connection a discussion of the role of the feedback term in our model is worth doing. In effect, the feedback term is serving two purposes - it breaks the symmetry and drives the system out of equilibrium. Actually, for $\beta>0$, 1. it reduces the damping experienced by the particle at $x_1$ and 2. it makes the temperature of the particle at $x_1$ different from that at $x_2$ (which is a non-equilibrium situation ensuring possibility of heat flow between the parts of the system). A change in the sign of $\beta$ would do the same but in a way so that the particle at $x_2$ would now be in the driver's seat by having smaller damping and larger temperature. The different damping experienced by two different particles could be because of two differently shaped bodies sitting at $x_1$ and $x_2$ and the difference in temperature could be due to the fact that one of them is being driven by some unbiased force. Within the present feedback picture, we do not need to take the shape of the objects at positions $x_1$ and $x_2$ into consideration, rather take them as particles. However, taking these two essential requirements like symmetry breaking and the driving in a correlated way through the velocity feedback is obviously a restriction. But, it nicely allows us to do a simple exact analytic calculation for the system to explicitly understand the interplay of the elasticity and volume exclusion to generate directional transport. An extension of the system in 2D, where rotational degrees of freedom are present, is straight forward \cite{ari} and it can generally capture the principle of transport of some bacterial systems.
 
{\bf Acknowledgement} I acknowledge stimulating discussions with Prof. Amos  Maritan.

\newpage
\begin{figure}
 \begin{psfrags} 
 \psfrag{v}[cc][][1.2][0]{$<\dot X>$}
 \psfrag{a}[cc][][1.5][0]{$\alpha$}
  \includegraphics[scale=0.5,angle=-90]{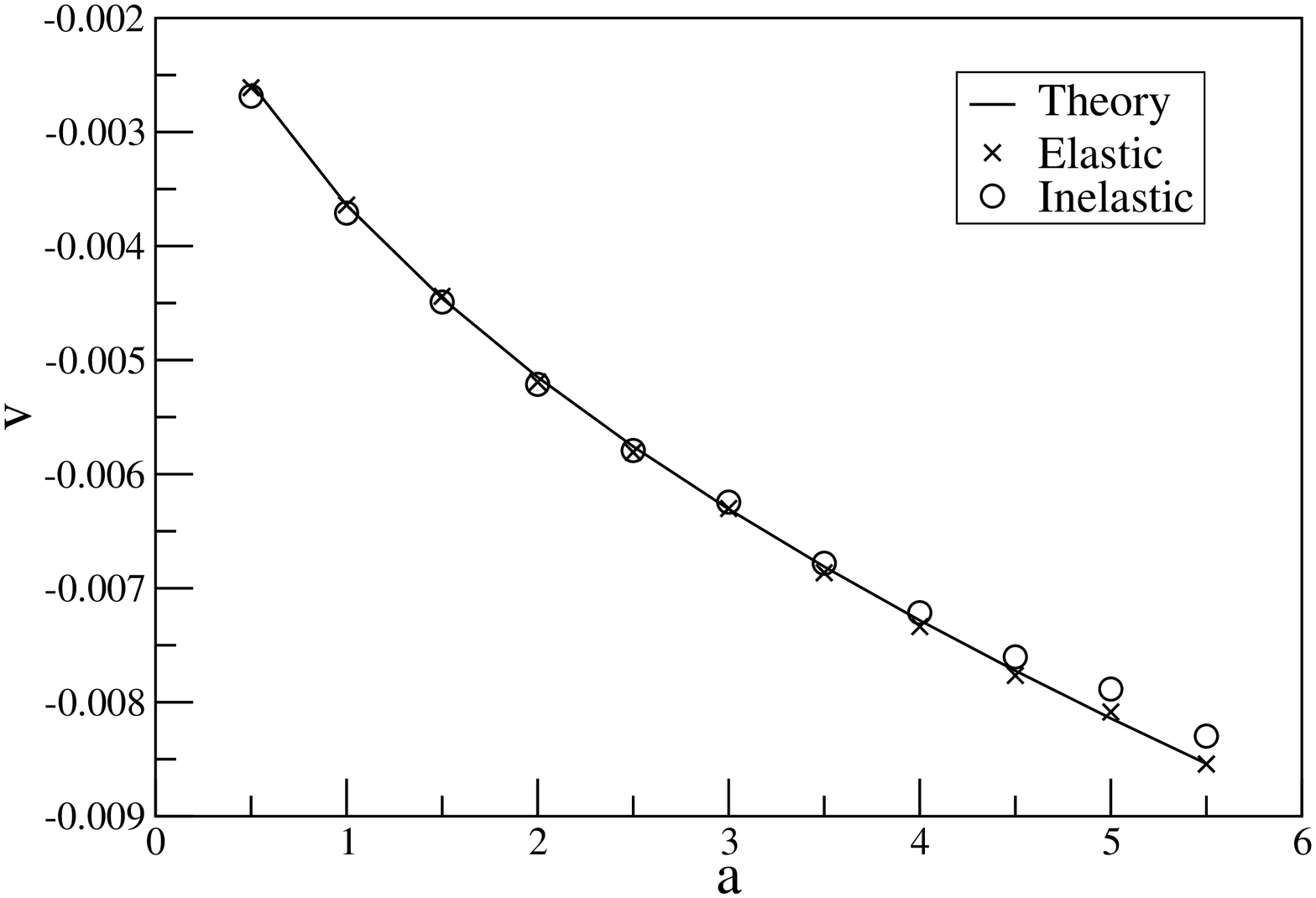}
 \end{psfrags}    
\textsf{\caption[Figure 1a]{ a. Average velocity of the CM vs $\alpha$.}}
\end{figure}

\addtocounter{figure}{-1}
\begin{figure}
 \begin{psfrags} 
 \psfrag{v}[cc][][1.5][0]{$<\dot X>$}
 \psfrag{b}[cc][][1.5][0]{$\beta$}
  \includegraphics[scale=0.5,angle=-90]{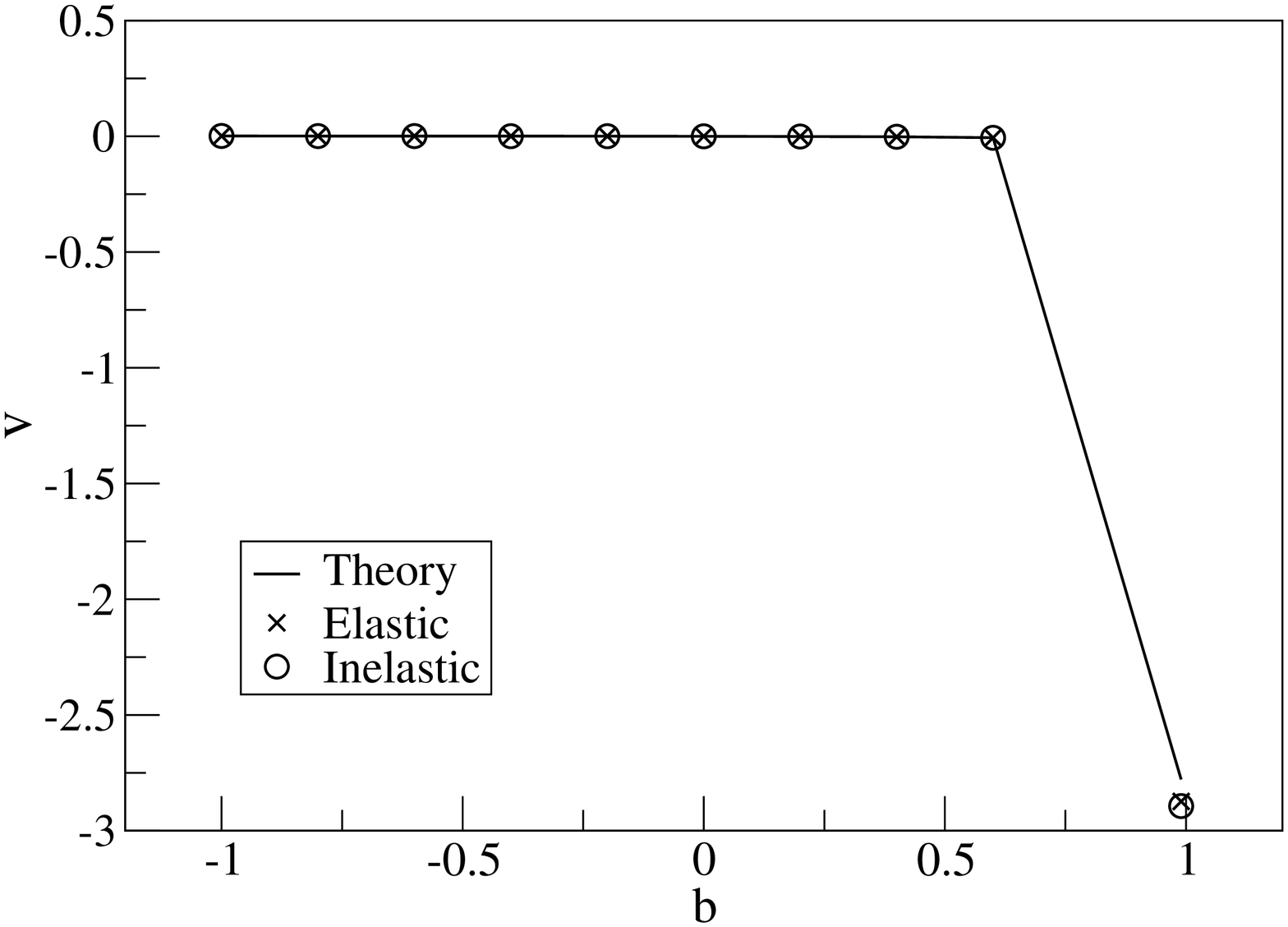}
 \end{psfrags}    
 \textsf{\caption[Figure 1b]{b. Average velocity of the CM vs $\beta$.}}
\end{figure}

\begin{figure}
 \begin{psfrags} 
 \psfrag{e}[cc][][1.5][-90]{$\eta$}
 \psfrag{a}[cc][][1.5][0]{$\alpha$}
  \includegraphics[scale=0.5,angle=-90]{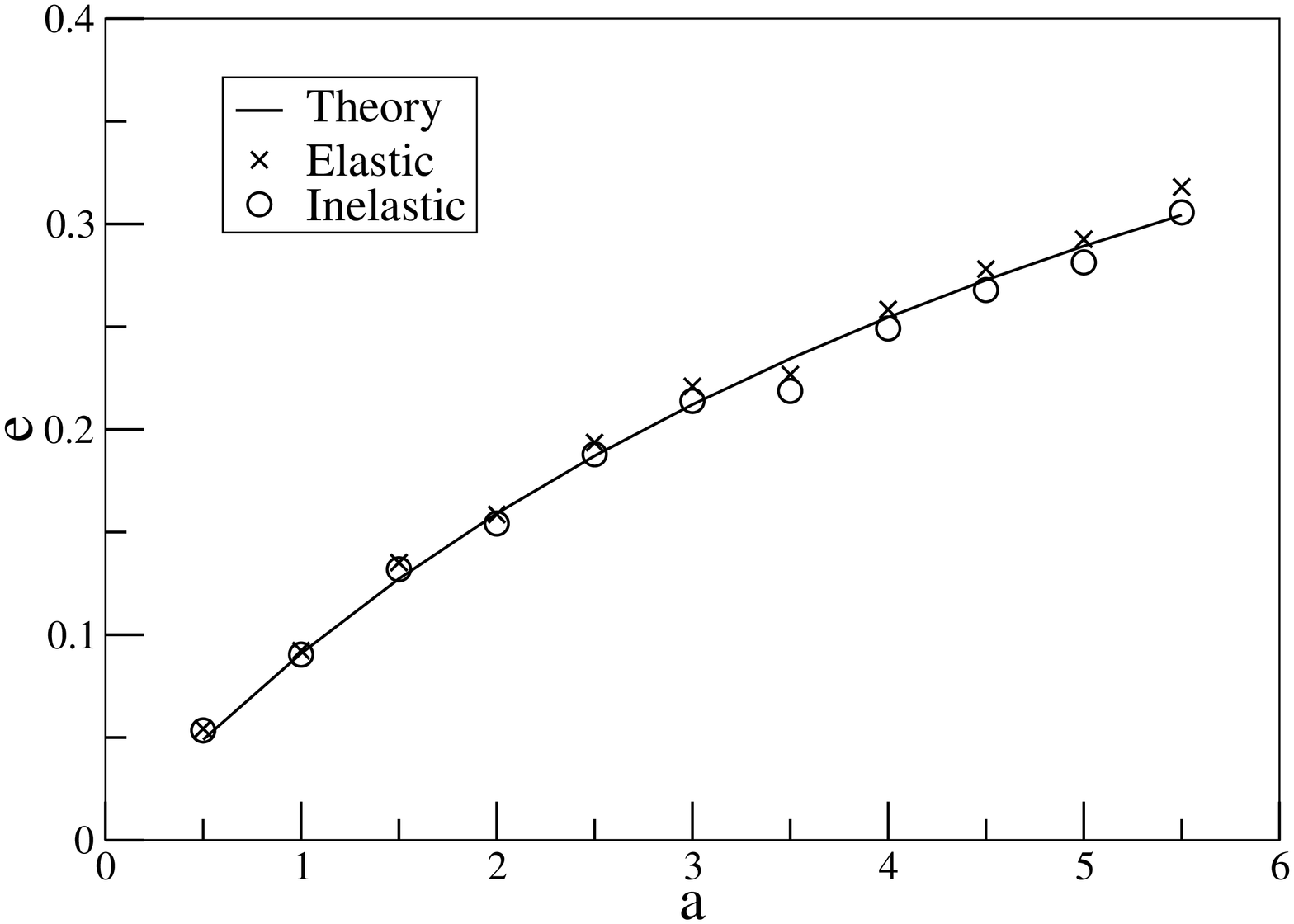}
 \end{psfrags}    
\textsf{\caption[Figure 2a]{a. Efficiency ($\eta$) vs $\alpha$.}}
\end{figure}

\addtocounter{figure}{-1}
\begin{figure}
 \begin{psfrags} 
 \psfrag{e}[cc][][1.5][-90]{$\eta$}
 \psfrag{b}[cc][][1.5][0]{$\beta$}
  \includegraphics[scale=0.5,angle=-90]{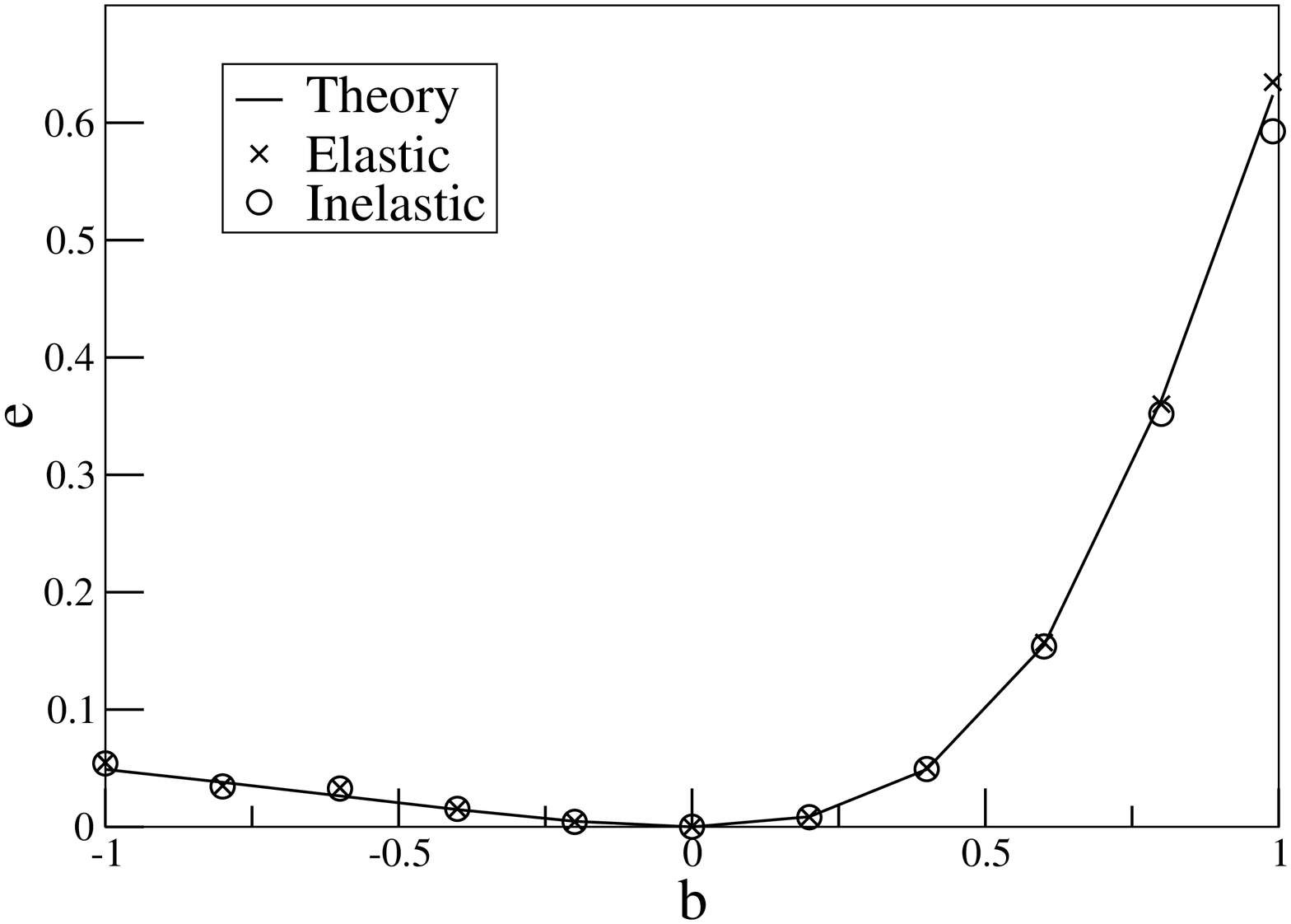}
 \end{psfrags}    
 \textsf{\caption[Figure 2b]{b. Efficiency ($\eta$) vs $\beta$.}}
\end{figure}


\newpage
\begin{thebibliography}
\noindent\bibitem {ast} RD Astumian, Science {\bf 276}, 917 (1997)
\noindent\bibitem {rei} P Reimann, Phys. Rep. {\bf 361}, 57 (2002) 
\noindent\bibitem {han} P Reimann and P H\"anggi, Appl. Phys. A {\bf 75}, 169 (2002)
\noindent\bibitem {han1} P H\"anggi and F Marchesoni, Rev. Mod. Phys., {\bf 81}, 387 (2009)
\noindent\bibitem {vale} A Yildiz, M Tomishige, RD Vale and PR Selvin, Science {\bf 303}, 676 (2004)
\noindent\bibitem {ast1} RD Astumian and I Der\'enyi, Biophys. J {\bf 77}, 993 (1999) 
\noindent\bibitem {oka} Y Okada and N Hirokawa, Science {\bf 283}, 1152 (1999)
\noindent\bibitem {gei} B Geislinger and R Kawai, Phys. Rev. E {\bf 74}, 011912 (2006)
\noindent\bibitem {jone} PH Jones, M Goonasekera and F Renzoni, Phys. Rev. Lett. {\bf 93}, 073904 (2004)
\noindent\bibitem {cil} S Cilla, F Falo and LM Flor\'{i}a, Phys. Rev. E {\bf 63}, 031110 (2001)
\noindent\bibitem {von} S von Gehlen, M Evstigneev and P Reimann, Phys. Rev. E {\bf 77}, 031136 (2008)
\noindent\bibitem {mogi} A Mogilner, M Mangel and RJ Baskin, Phys. Lett A {\bf 237}, (1998)
\noindent\bibitem {fog} HC Fogedby, R Metzler and A Svane, Phys Rev. E {\bf 70}, 021905 (2004)
\noindent\bibitem {nor} B Nord\'{e}n, Y Zolotaryuk, PL Christiansen and AV Zolotaryuk, Phys. Rev. E {\bf 65}, 011110 (2001)
\noindent\bibitem {kum} KV Kumar, S Ramaswamy and M Rao, Phys. Rev. E {\bf 77}, 020102(R) (2008)
\noindent\bibitem {bro} M van den Broek, R Eichhorn and C van den Broeck, EPL {\bf 86}, 30002 (2009)
\noindent\bibitem {ari} A. Bhattacharyay {\bf arXiv:0905.2294}

\end{thebibliography}
\end{document}